\documentclass[journal=jacsat,manuscript=article,layout=twocolumn]{achemso}

\usepackage{chemformula} % Formula subscripts using \ch{}
\usepackage[T1]{fontenc} % Use modern font encodings
\usepackage[colorlinks,linkcolor=blue, urlcolor=blue, anchorcolor=blue, citecolor=blue]{hyperref}
\usepackage{diagbox}

\author{Yuanjin Wang}
\affiliation[Beijing Institute of Technology]
{Center for Quantum Technology Research, Key Laboratory of Advanced Optoelectronic Quantum Architecture and Measurements (MOE), School of Physics, Beijing Institute of Technology, Beijing, China}
\author{Gaoshang Li}
\affiliation[Beijing Institute of Technology]
{Center for Quantum Technology Research, Key Laboratory of Advanced Optoelectronic Quantum Architecture and Measurements (MOE), School of Physics, Beijing Institute of Technology, Beijing, China}
\author{Jin Dai}
\affiliation[Beijing Institute of Technology]
{Center for Quantum Technology Research, Key Laboratory of Advanced Optoelectronic Quantum Architecture and Measurements (MOE), School of Physics, Beijing Institute of Technology, Beijing, China}
\email{jindai@bit.edu.cn}
\author{Xubiao Peng}
\affiliation[Beijing Institute of Technology]
{Center for Quantum Technology Research, Key Laboratory of Advanced Optoelectronic Quantum Architecture and Measurements (MOE), School of Physics, Beijing Institute of Technology, Beijing, China}
\email{xubiaopeng@bit.edu.cn}
\author{Qing Zhao}
\affiliation[Beijing Institute of Technology]
{Center for Quantum Technology Research, Key Laboratory of Advanced Optoelectronic Quantum Architecture and Measurements (MOE), School of Physics, Beijing Institute of Technology, Beijing, China}
\alsoaffiliation{Beijing Academy of Quantum Information Sciences, Beijing 100193, China}
\email{qzhaoyuping@bit.edu.cn}

\title[An \textsf{achemso} demo]
  {Research on the transition dynamics and linear (nonlinear) optical properties of mCherry}

\abbreviations{IR,NMR,UV}

\begin{document}

%%%%%%%%%%%%%%%%%%%%%%%%%%%%%%%%%%%%%%%%%%%%%%%%%%%%%%%%%%%%%%%%%%%%%
%% The "tocentry" environment can be used to create an entry for the
%% graphical table of contents. It is given here as some journals
%% require that it is printed as part of the abstract page. It will
%% be automatically moved as appropriate.
%%%%%%%%%%%%%%%%%%%%%%%%%%%%%%%%%%%%%%%%%%%%%%%%%%%%%%%%%%%%%%%%%%%%%

%%%%%%%%%%%%%%%%%%%%%%%%%%%%%%%%%%%%%%%%%%%%%%%%%%%%%%%%%%%%%%%%%%%%%
%% The abstract environment will automatically gobble the contents
%% if an abstract is not used by the target journal.
%%%%%%%%%%%%%%%%%%%%%%%%%%%%%%%%%%%%%%%%%%%%%%%%%%%%%%%%%%%%%%%%%%%%%
\begin{abstract}
In this study, we explore the electron transition mechanism and optical properties of the popular red fluorescent protein mCherry. By examining the charge transfer spectrum and combining it with the mCherry hole-electron distribution, we identify that the charge transfer between the phenolate and imidazolinone loops significantly contributes to the absorption spectrum. Quantitative analysis of charge transfer shows that, overall, the electrons are transferred to the $C16$ atom in the middle of phenolate and the imidazolinone loops during absorption. We speculate that $C16$ may also absorb protons to enable the photoconversion of mCherry in the excited state, similar to the blinking mechanism of IrisFP. In addition, we further investigated the optical properties of mcherry in the external field by polarizability (hyperpolarizability), showing the anisotropy of the polarization, the first hyperpolarization and the second hyperpolarization by unit spherical representation. Our results suggest that significant polarization and second hyperpolarizability occur when the field direction and electron transfer direction are aligned. We also analyzed the polarizability and first hyperpolarizabilities for different external fields. The polarizability mutated when the external field satisfies the $S_{0,min}\rightarrow S_{1}$ transition. Finally, the study of the first hyperpolarizability shows that adjusting the appropriate field can lead to a linear photoelectric effect or second harmonic generation of mCherry. These studies have certain reference values for various red fluorescent protein correlation simulations and experiments because of the similarity of the red fluorescent protein.

\end{abstract}

%%%%%%%%%%%%%%%%%%%%%%%%%%%%%%%%%%%%%%%%%%%%%%%%%%%%%%%%%%%%%%%%%%%%%
%% Start the main part of the manuscript here.
%%%%%%%%%%%%%%%%%%%%%%%%%%%%%%%%%%%%%%%%%%%%%%%%%%%%%%%%%%%%%%%%%%%%%
\section{Introduction}
\begin{figure*}
  \includegraphics[width=0.95\linewidth]{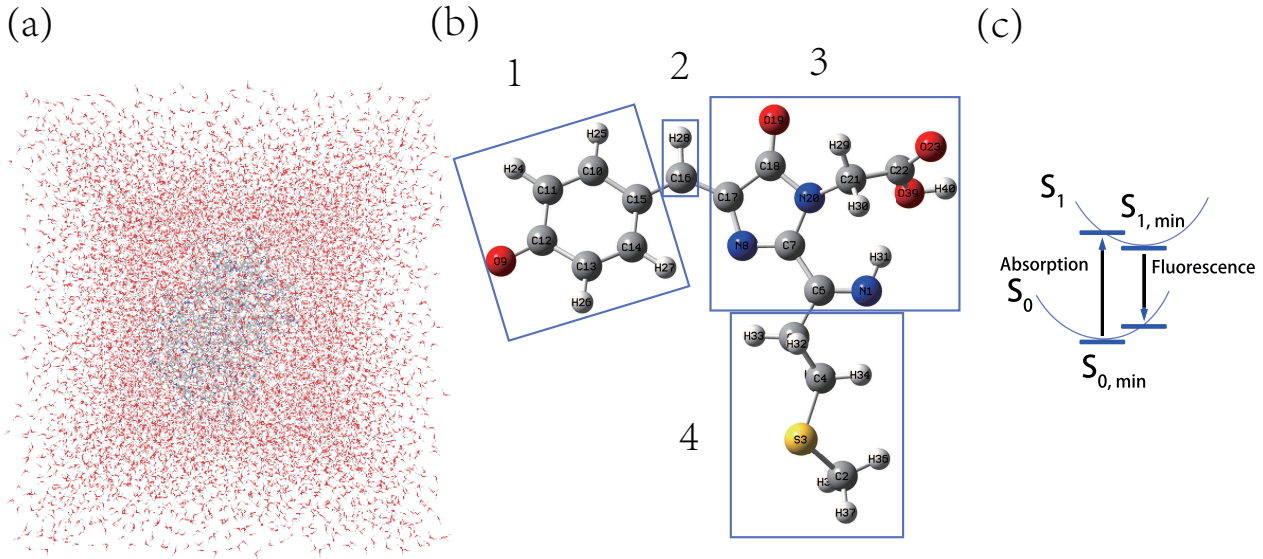}
  \caption{(a) Crystal structure of mCherry. Protein in the middle and water molecules in the periphery. (b) Chromophere. Gray, red, white, blue, and yellow colors refer to the $C$, $O$, $H$, $N$ and $S$ atoms, respectively. The chromophore was divided into four regions for analysis. (c) Absorption and fluorescence process diagram. $S_{0}$ means that the molecule is in the ground state, and $S_{0,min}$ is the minimum point of the ground state. The first excited state and its minimum are $S_{1}$ and $S_{1,min}$, respectively. $S_{0,min}\rightarrow S_{1}$ corresponds to the absorption process, while $S_{1,min}\rightarrow S_{0}$ is the fluorescence process. }
  \label{fig1}
\end{figure*}
Fluorescent proteins (FPs), derived from green fluorescent proteins found in the jellyfish \emph{Aequorea victoria}\cite{MISHIN20151}, can be engineered to respond to various biological events and signals, rendering them valuable for various biological tissues and organisms while rarely causing photodynamic toxicity\cite{Zhang2002}.

Decades of research have resulted in a family of FPs that span the entire visible spectrum, leading to the development of powerful fluorescent probes and revealing the complexity and application value of the photophysical properties of proteins\cite{doi:10.1021/acs.jpcb.1c05629,Dean2014,MonteroLlopis2021,Koch2018,BANDO2019802,doi:10.1021/acs.jafc.8b04646,Ranjit2021}. In bioimaging and sensing, FPs and their derived biosensors make it possible to probe the location, activities, or interactions of molecules from the subcellular to multicellular scales\cite{DUWE2019183}. Fluorescence imaging technology is ideal for measuring the pathophysiological microenvironment measurement\cite{D1CS00083G}, and fluorescent protein nanothermometers have the capability of accurately targeting to organelles and genetically encode\cite{Zhou2020}. Tumour-targeted fluorescence-guided surgery enables the visualization of solid tumors\cite{Mieog2022}, while fluorescence lifetime imaging and F$\ddot{o}$rster resonance energy transfer has been coupled with multiphoton microscopy for in vivo dynamic imaging\cite{Coelho:20}. Many photoinduced reactions of FPs, such as photoisomerization, excited-state proton transfer, photooxidation/photoreduction, are initiated by electron transfer\cite{doi:10.1021/acs.chemrev.6b00238}. The reversible photoconversion characteristic of fluorescent proteins to switch back and forth between fluorescent and nonfluorescent states is highly significant to the development of advanced fluorescence imaging and biotechnology based on their photophysical properties\cite{Coquelle2018,adam2014phototransformable,C3CS60171D}. Engineering FPs to switch between dark and detectable states enables to image structures with spatial resolution beyond Abbe's diffraction limit\cite{doi:10.1126/science.aau1044}. Dynamic fluorescence can adjust surface patterns through external stimulation, which is highly valuable in the fields of smart displays, information storage, and anticounterfeiting\cite{Ma2020,doi:10.1021/jacs.7b07738,doi:10.1021/acsami.8b14110,Gao2018,Liu2019}.

Red fluorescent proteins (RFPs) have low scattering and large penetration depth in the red to near-infrared region,  making them advantageous in deep-tissue imaging of live animals owing to their longer excitation and emission wavelengths\cite{doi:10.1021/acs.jpcb.1c05629,doi:10.1021/acs.analchem.9b00224}. For instance, the bright red fluorophore mRuby3 can be used to design a voltage-activated red neuronal activity monitor, extending the application of genetically encoded voltage indicators in high-speed multispectral imaging and opening the door to high-speed multispectral functional imaging in vivo\cite{kannan2018fast}. The energy of the excited state of RFPs strongly correlates with the magnetic dipole moment, which can be used to adjust the spectrum of markers in in vivo imaging\cite{doi:10.1021/acs.jcim.1c00981}.

Although mCherry has low fluorescence quantum yield, it is a popular red fluorescent protein, the lineage of which can be traced to the naturally occurring tetramer DsRed\cite{Shaner2004}. However, naturally occurring RFPs occur in dimeric or tetrameric form, both of which tend to oligomerize and are unsuitable for fusion tagging\cite{Matz1999,doi:10.1073/pnas.98.2.462,doi:10.1073/pnas.082243699,Merzlyak2007}. Several novel mutant monomeric variants known as mFruits have been designed to address this issue; among the designed variants, mCherry, mOrange, and mStrawberry are the most promising\cite{Shaner2004,Shaner2005,doi:10.1063/1.3660197}. Compared to its progenitor, mCherry has the advantages of a favorable red-shift of absorption and emission spectra, higher expression and fast chromophore maturation, and lower phototoxicity\cite{mukherjee_manna_hung_vietmeyer_friis_palmer_jimenez_2022}. Unfortunately, the fluorescence quantum yield of mcherry is only one-third, rendering it suubstantially dimmer than DsRed\cite{Shaner2004,doi:10.1073/pnas.97.22.11984}. The second-order nonlinear optical response ($\beta$) of fluorescent proteins including mCherry can be measured via Hyper-Rayleigh Scattering and is suitable for second-harmonic imaging microscopy\cite{doi:10.1021/ja400098b}. RFPs with a common origin are similar, and their transition mechanism in absorption and emission is intriguing. However, related studies on this photoinduced reaction are relatively few. Consequently, we posit that studying the transition mechanism of mCherry in detail could provide some references for experiments related to mCherry and other RFPs.

In this study, we used the QM/MM method to simulate the absorption process of mcherry, and we analyzed the calculated wave function using Multiwfn (Version $3.8$(dev))\cite{https://doi.org/10.1002/jcc.22885} software to investigate the transition mechanism and optical properties of mCherry. The absorption process involves molecular transition from the ground state minimum ($S_{0,min}$) to the excited state ($S_{i}, i=1,2,3 \ldots$); we consider only the $S_{0,min}\rightarrow S_{1}$ transition with the largest contribution to the mCherry absorption spectrum. Using Multiwfn, we introduced hole-electron analysis to study the transition mechanism of mCherry. First, we observe the hole-electron distribution \cite{LIU2020468} of the excited state, and combined with the charge transfer spectrum\cite{LIU202278} of mCherry, we investigate the influence of the charge transfer within the chromophore on the absorption spectrum. Second, we quantitatively analyzed charge transfer during absorption using the interfragment charge transfer module\cite{Multiwfn3.21.8} of Multiwfn and plotted a heat map. Additionally, we focus on the palarization and hyperpolarization calculated by the complete state summation method\cite{doi:10.1063/1.466123}, showing the polarizability, the first and second hyperpolarizations in all directions using the unit spherical representation\cite{https://doi.org/10.1002/jcc.21694,D1CP05883E}. Finally, we studied the polarizability and the first hyperpolarization under different external fields and analyzed the linear and nonlinear optical properties of mCherry.

\section{Models and methods}
Herein, we used ORCA (Version $5.0.3$)\cite{https://doi.org/10.1002/wcms.81} software for QM/MM calculation and Multiwfn\cite{https://doi.org/10.1002/jcc.22885} software for data analysis. To prepare for the analysis, we obtained the mCherry crystal structure file (code 2H5Q)\cite{doi:10.1021/bi060773l} from the Protein Data Bank. Then, we used VMD\cite{HUMPHREY199633} software to cover it with a layer of water molecules $15$\AA $ $ thick (Fig.\hyperref[fig1]{1(a)}), and generate a protein structure file (PSF), in which residue glu215 was adjusted to a protonation state.

Using Avogadro\cite{Hanwell2012}, we individually added $H$ atoms to the chromophore and additionally appended $OH^{-}$ and $H^{+}$ ions at the interface to compensate for the effects of bond breakage. As shown in Fig.\hyperref[fig1]{1(b)}, the three extra attached atoms are $O39$, $H40$, and $H31$ at the right end. To investigate the role of phenolate and imidazolinone, we referred to the three-state adiabatic model\cite{doi:10.1063/1.3121324,D1SC05849E} to divide the chromophore into four regions. Region $1$ comprises the phenolate ring, Region $3$ includes the imidazolinone ring and partially surrounding atoms, Region $2$ comprises the phenolate ring and the imidazolinone ring middle region, and the remaining parts are included in Region $4$. The PSF file was then transformed into an ORCA force field file, and the force field file of the chromophore was generated directly in the ORCA software using the \textbf{orca\_mm} module. Subsequently, the mCherry was pre-optimized using the L-BFGS optimizer (keyword “L-Opt”) in ORCA.

To facilitate the analysis, we performed a geometric manipulation of the molecule using Multiwfn. In order, making the $C-H$ bond of region $2$ parallel to the $Y$-axis, selecting the heavy atom center of region $1-3$ as the origin of coordinates, and adjusting the phenolate ring of region $1$ parallel to the $x-y$ plane.

Next, we initiated the QM/MM calculation using ORCA software. As shown in Fig.\hyperref[fig1]{1(c)}, we regard $S_{0,min}\rightarrow S_{1}$ directly as the absorption process, while the fluorescence process is $S_{1,min}\rightarrow S_{0}$. The structure optimization level was CAM-B3LYP(D3)/cc-pVDZ, where the QM region contained only the chromophore, and the active region included all atoms within $12$\AA $ $ of the chromophore. We then selected the higher level DSD-PBEP86(D3)/may-cc-pV(T+d)Z to perform TDDFT calculations and imported the results into Multiwfn for spectra drawing and various analyses.

We analyze the linear (nonlinear) optical characteristics of mCherry in different external fields by polarization (hyperpolarization). In Multiwfn, the complete state summation method was used to calculate polarization and hyperpolarizability\cite{doi:10.1063/1.466123}. The Taylor expansion of the system energy $E$ relative to the uniform external field $F$ is given as\cite{Multiwfn}

\begin{equation*}
E(F) = E(0)-\mu_0 F-\frac{1}{2} \alpha F^2-\frac{1}{6} \beta F^3-\frac{1}{24} \gamma F^4\ldots.
\end{equation*}
where $\mu_0$ is called the permanent dipole moment, $\alpha$ is the polarizability (linear optical coefficient), $\beta$ is the first hyperpolarizability (second-order nonlinear optical coefficient) and $\gamma$ is the  second hyperpolarizability (third-order nonlinear optical coefficients) , which satisfies
\begin{footnotesize}
$$
\begin{aligned}
\mu_0=-\left.\frac{\partial E}{\partial F}\right|_{F=0},\alpha=-\left.\frac{\partial^2 E}{\partial F^2}\right|_{F=0},\\
 \beta=-\left.\frac{\partial^3 E}{\partial F^3}\right|_{F=0},\gamma=-\left.\frac{\partial^4 E}{\partial F^4}\right|_{F=0}.
\end{aligned}
$$
\end{footnotesize}

In addition, the total dipole moment $\mu=-\frac{\partial E}{\partial F}$ can be expressed as
\begin{equation*}
\mu=\mu_0+\underbrace{\alpha F}_{\mu_1}+\underbrace{(1 / 2) \beta F^2}_{\mu_2}+\underbrace{(1 / 6) \gamma F^3}_{\mu_3}+\ldots
\end{equation*}
where $\mu_i (i=1,2,3)$ is called the induced dipole moment. Assuming that the external field per unit strength is used, the polarizability (hyperpolarizability) under the electrostatic field can measure the change in the molecule-induced dipole moment. Generally, molecules with stronger polarizability tend to have larger molecular volumes\cite{brink1993murray,https://doi.org/10.1002/asia.202100589}.

\section{Results and discussion}
\begin{figure}
  \includegraphics[width=\linewidth]{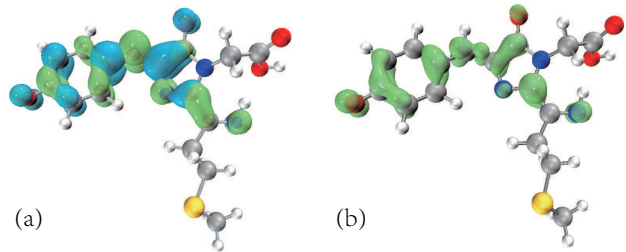}
  \caption{(a) Hole-electron distribution in the $S_{1}$ state. The blue surface represents the hole distribution, whereas the green surface represents the electron distribution. (b) Function $Sr$ distribution diagram. The smooth green surface reflects the hole-electron overlap region.}
  \label{fig2}
\end{figure}
\begin{figure}
  \includegraphics[width=\linewidth]{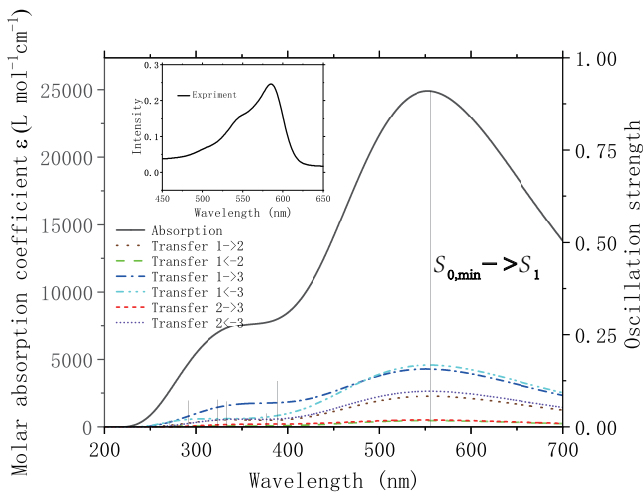}
  \caption{Charge-transfer spectrum. The solid black solid line is the normal absorption spectrum, and the dots or lines of other colors indicate the contribution to the absorption spectrum when electrons transfer between regions $1-3$. Vertical lines correspond to the right axis, indicating the oscillator strength. The highest one vertical line corresponds to the $S_{0,min}\rightarrow S_{1}$ transition. The subgragh shows the absorption spectrum of mCherry measured in our experiment.}
  \label{fig3}
\end{figure}
\begin{figure*}
  \includegraphics[width=\linewidth]{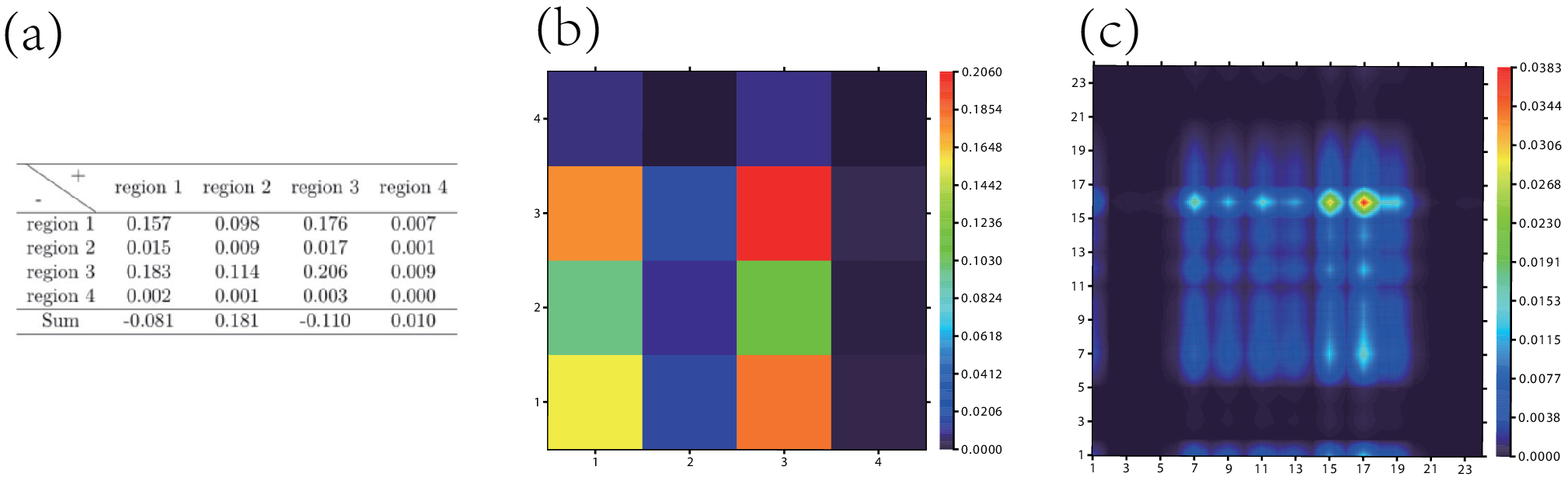}
  \caption{(a) Charge transfer between regions during $S_{0,min}\rightarrow S_{1}$; $-(+)$ represents the region where electrons are lost (obtained), and the last row represents the total change in the number of electrons. (b) The corresponding heat map. The horizontal (vertical) coordinates indicate the region of lost (obtained) electrons. (c) Heat map showing the charge transfer between heavy atoms. The horizontal (vertical) coordinates represent heavy atoms that have lost (obtained) electrons. The redder the color in (b) and (c), the more electrons are transferred.
}
  \label{fig4}
\end{figure*}

\textbf{Hole-Electron Distribution.} Fig.\hyperref[fig2]{2(a)} shows the hole-electron distribution in the $S_{1}$ state of mCherry. It is apparent that the holes and electrons are mainly distributed around the phenolate ring and the imidazolinone ring. Based on Fig.\hyperref[fig1]{1(b)}, electron transfer exclusively occurs in Regions $1-3$ and no hole-electron distribution in Region $4$. In the absorption process, the molecule transitions from the ground state to the first excited state, loses electrons in the region of the blue grid surface and obtains electrons in the region of the green grid surface, eventually leading to massive electron transfer inside Regions $1-3$. In addition, we have attempted to expand the QM region, but the final hole-electron distribution is not significantly different.

Fig.\hyperref[fig2]{2(b)} indicates the region where the electrons and holes significantly overlap. As can be clearly observed therein, the electrons and holes overlapped very heavily. Moreover, based on the output of Multiwfn, the centroid distance between the hole and the electron is only $0.405$\AA, which is less than half of the $C-C$ single bond (about $1.55$ \AA \cite{HAO2018222}). From the high overlap of holes and electrons, we infer that the $S_{0,min}\rightarrow S_{1}$ transition is accompanied by a large transition electric dipole moment, corresponding to a large oscillator intensity, eventually producing a strong absorption peak in the absorption spectrum\cite{doi:10.1119/1.12937}.

\begin{figure*}
  \includegraphics[width=\linewidth]{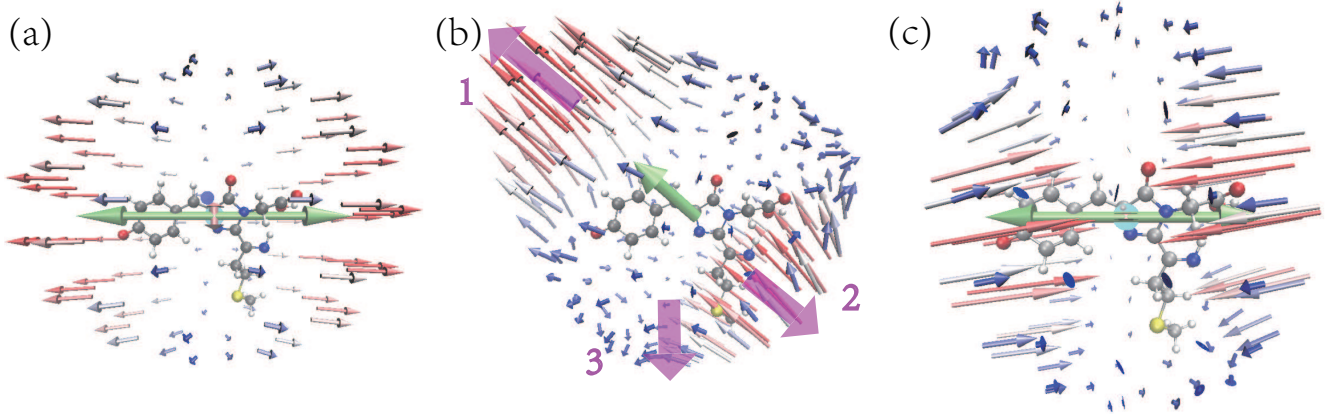}
  \caption{The polarization (a), first hyperpolarization (b), and second hyperpolarization (c) shown in unit spherical representation. The longer the arrow is, the redder it is, indicating that if outfields with unit strength are added from the molecular center to the direction, the greater the polarization, the first hyperpolarization and the second hyperpolarization will be. The shorter the arrow, the bluer it is, indicating the opposite meaning. In (a) and (c), the green double arrow, the orange double arrow (center) and the blue circle (center) represent  the components in the $x$, $y$, $z$ direction, respectively. In (b), the green arrow represents the overall orientation of the first hyperpolarizability, and the pink arrows are added for convenience. }
  \label{fig5}     
\end{figure*}

\textbf{Charge-Transfer Spectrum.} The absorption spectrum of the mCherry and the contribution of charge transfer between regions to the absorption spectrum are shown in Fig.\ref{fig3}. According to the hole-electron distribution in Fig.\hyperref[fig2]{2(a)}, there are no holes or electrons in Region $4$; hence, we only consider charge transfer between Regions $1-3$. The dark blue dot-dash line and the light blue double dot-dash line have the largest effect on the spectrum, representing the charge transfer between Regions $1$ and $3$, namely between the phenolate and imidazolinone rings. The brown dotted line and the purple short dotted line have a slightly less impact, representing the charge transfer from Region $1$ and Region $3$ to Region $2$, i.e., the contribution to the absorption spectrum when the charges in the phenolate and imidazolinone rings move to the region in between them. Fig.\ref{fig3} clearly shows that the oscillator intensity corresponding to the $S_{0,min}\rightarrow S_{1}$ transition is the largest, and the contribution to the absorption spectrum is the largest, confirming our speculation. The subfigure shows the mCherry absorption spectrum measured in our experimentally, and its trend is consistent with our calculation results. Combined with the consistent trend of the experimental and calculated results and the above conclusion that the hole-electron distribution does not change with the increase in the QM region, we conclude that our study can describe the mCherry transition mechanism.

\textbf{Interfragment Charge Transfer.}  Fig.\hyperref[fig4]{4} shows the specific electron transfer values between regions or heavy atoms. Combining Fig.\hyperref[fig3]{3} and \hyperref[fig4]{4(a)}, it can be found that many electrons are transferred between Region 1 and Region 3, which also makes great contributions to the absorption spectrum. The more the number of electrons transferred, the greater the contribution to the absorption spectrum. With $1.157$ electrons transferred within Region $1$ and simultaneously $0.206$ electrons moving within Region $3$, it is conceivable that both parts also substantially contribute to the absorption spectrum. Overall, Region $1$ lost $0.081$ electrons, Region $3$ lost $0.11$ electrons, and instead, Region $2$ obtained $0.181$ electrons. Therefore, electrons generally move from phenolate and imidazolinone to the $C16$ atom between them. This conclusion is different from that of electron transfer obtained from observing the positive (negative) electron density difference centroid, in the case of which the electrons move purely from phenolate to imidazolinone\cite{doi:10.1021/acs.jcim.1c00981}. The hole-electron distribution map directly shows the original distribution characteristics of holes and electrons without cancelation, which has certain advantages in investigating the intrinsic characteristics of electron excitation. Furthermore, if we judge the direction of electron transfer by referring to the hole centroid and the electron centroid, the result is consistent with that observed from the perspective of density difference.

The electron transfer between regions can be more intuitively understood from Fig.\hyperref[fig4]{4(b)}, in which a large number of electrons are transferred into Region $2$ (colored green), whereas very few atoms are transferred out of Region $2$ (colored dark blue). The number of electrons transferred between Regions $1$ and $3$ is very close, both shown in orange. Region $3$ has the highest number of redistributed electrons, shown in red. Additionally, if the four regions are not divided, the heat mat of the direct study on charge transfer between heavy atoms, as shown in Fig.\hyperref[fig4]{4(c)}, can be found from another perspective that electrons are concentrated in $C16$ in Region $2$, and the nearby $C15$ and $C17$ transfer an extra large number of electrons to $C16$. In contrast, the result of electron transfer is that $C16$ is negatively charged, with the possibility of proton absorption, which is very similar to the blinking mechanism\cite{doi:10.1021/ja2085355} in IrisFP. The blinking mechanism\cite{ADAM201492} of IrisFP is the thermally reversible photoinduced proton exchange between the methylene bridge of the chromophore and the surrounding residues. In particular, the chromophore of IrisFP acquired protons at exactly the same site as that of $C16$ in the chromophore of mCherry. Therefore, it is possible for $C16$ to absorb the proton and protonize the chromophore, resulting in phototransformation of mCherry.

\textbf{Linear and Nonlinear Optical Properties.} Fig.\hyperref[fig5]{5} displays the anisotropy of polarization, the first hyperpolarization, and the second hyperpolarization using unit spherical representation. In Fig.\hyperref[fig5]{5(a)} and Fig.\hyperref[fig5]{5(c)}, the components in the $y$ and $z$ directions are negligible relative to the $x$ direction. Thus, applying the external fields in the connection direction of the phenolate ring and the imidazolinone ring can generate relatively large polarization and second hyperpolarization. In contrast, the polarization and second hyperpolarization are relatively negligible when the outer fields are perpendicular to the $x$ direction. According to the electron transfer situation shown in Fig.\hyperref[fig4]{4}, the external field applied in $x$ direction is in the same direction as the electron transfer, resulting in the largest influence on polarization or hyperpolarization. In contrast, the influence of the external field perpendicular to the electron transfer direction is negligible. As for Fig.\hyperref[fig5]{5(b)}, the first hyperpolarization generated by outfields applied in direction $1$ is larger and in the same direction as the outfields, but the first hyperpolarization generated by outfields applied in direction $2$ is opposite to the outfields, whereas the first hyperpolarization generated by outfields applied in direction $3$ is smaller and perpendicular to the outfields.

\begin{figure}
  \includegraphics[width=\linewidth]{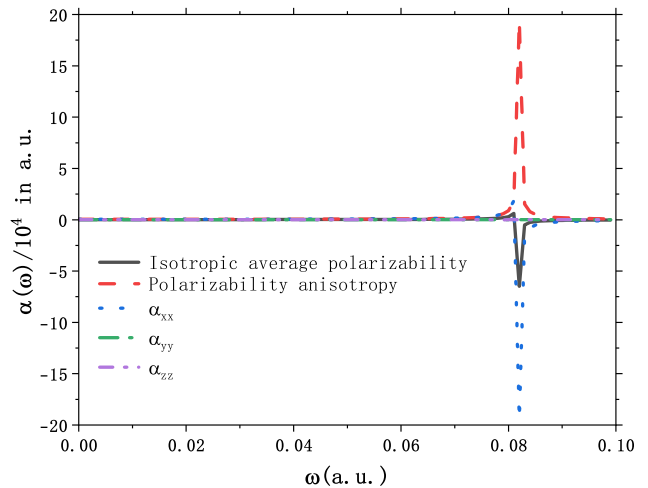}
  \caption{Polarizability ($\alpha$) in different external fields (frequency $\omega$). The solid black line is the isotropic average polarization, and the red dashed line is the polarizability anisotropy. The blue dot line, the green dot-dash line, and the purple double dotted line represent the polarization components in the $x$, $y$, and $z$ directions, respectively.}
  \label{fig6}
\end{figure}
	Fig.\hyperref[fig6]{6} studies the variation of the polarizability with the frequency of the outfield. It shows that when the external field frequency is $0.082$ a.u. (about $556$ nm), satisfying the $S_{0,min}\rightarrow S_{1}$ transition, the polarizability is mutated while the polarizabilities in all directions around the transition point obviously differ from each other. Specifically, the main reason is that the polarization of $x$ direction undergoes a huge change while the polarizations of the $y$ and $z$ directions can be ignored in terms of both the absolute size and the degree of change. Moreover, the absolute value of anisotropic polarization (orange) is consierably higher than that of isotropic average polarization (black), and the polarization of mCherry presents significant anisotropy.

\begin{figure}
  \includegraphics[width=\linewidth]{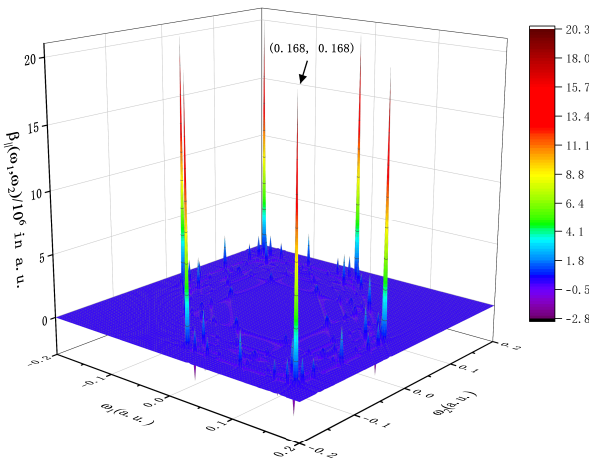}
  \caption{A scan of the first hyperpolarizability in the direction parallel to the dipole moment ($\boldsymbol{\beta}_{\|}$) as a function of  the two external field frequencies ($\omega_{1}, \omega_{2}$). The color ranges from purple to red; the higher the second hyperpolarization, the redder the color.}
  \label{fig7}
\end{figure}
The scanning of the first hyperpolarizability as a function of the frequency of the field in Fig.\hyperref[fig7]{7} demonstrates  the excellent nonlinear optical characteristics of mCherry. The positive and negative signs of frequencies $\omega_{1}$ and $\omega_{2}$ have no physical significance. If the peak occurs when $\omega_{1}$ and $\omega_{2}$ symbols are the same, it corresponds to the combined frequency effect. In contrast, when the peaks with the opposite $\omega_{1}$ and $\omega_{2}$ symbols reflect the difference frequency effect. Specifically, the peak at $\omega_{1}=\omega_{2}$ represents the second harmonic generation effect at the corresponding frequency; in the absence of any adjustment in these two fields, if the effect is optical rectification, the peak appears at $\omega_{1}=-\omega_{2}$\cite{doi:10.1021/acsanm.9b00089,Multiwfn4.200.8.1}.

Abundant peaks appeared in our simulation, with six peaks far exceeding the others in height. Four of the six highest peaks correspond to $\omega_{1}$ or $\omega_{2}$ values of $0$, reflecting a linear photoelectric effect, and can be used in fields such as electric-optical regulators. Two of the highest peaks satisfy $\omega_{1}=\omega_{2}$ and generate second harmonics at their locations, which can be used in frequency multipliers and second-harmonic generation microscopy\cite{doi:10.1021/ja400098b,He:22}. Observing the external field frequencies corresponding to these six peaks, if the frequency of one field is $0$, the other external field frequency size is $0.168$ a.u.($271.2$ nm); in contrast, when $\omega_{1}=\omega_{2}$, both external field frequency sizes are $0.168$ a.u. Using different calculation methods, the frequency and wavelength of the nonzero field at the peak will change but the trend of the first hyperpolarizability changing with outfields is similar.

\section{CONCLUSIONS}
Here, we investigate the transition mechanism of mCherry during the absorption process and its linear (nonlinear) optical properties. With Multiwfn software, we introduced the hole-electron analysis method to investigate the transition mechanism of mCherry. Based on the absorption spectra we simulated with the QMMM method is consistent with the trend we obtained experimentally, and that expanding the QM region does not affect the hole-electron distribution, i.e., the electron transfer is concentrated in the chromophore, we judge our simulation and analysis are reliable. According to the charge transfer spectrum, a substantial charge transfer occurred between phenolate and imidazolinone, which considerably contributed to the absorption spectrum. Interestingly, the electrons converge overall towards the $C16$ atom between the phenolate ring and the imidazolinone ring during absorption, which different from the observed direction of electron movement along phenolate to imidazolinone from the perspective of the density difference\cite{doi:10.1021/acs.jcim.1c00981}. However, from the perspectives of the hole centroid and the electron centroid, the electron transfer direction was the same as the density difference perspective. Similar to the blinking mechanism\cite{ADAM201492,doi:10.1021/ja2085355} of IrisFP, the chromophore of mCherry may also absorb proton at $C16$ after excitation, thus achieving protonation and resulting photoconversion of mCherry.

In addition to investigating the transition mechanism of mCherry during the absorption process, we further analyzed its linear and nonlinear optical properties in different external fields. We utilized the unit spherical representation method to study the anisotropy of polarization, first hyperpolarization, and second hyperpolarization. Our results show that the polarizability and second hyperpolarizability are considerably larger in the phenolate and imidazolinone connection direction than when perpendicular to this direction. In connection with the previous analysis of the transition mechanism, this phenomenon is due to electrons moving mainly in this direction. Then, we study the polarizability in the outer field at different frequencies, and find that the polarizability is mutated in the connection direction of the phenolate loop and the imidazolinone loop when the outer field satisfies the $S_{0,min}\rightarrow S_{1}$ transition. Finally, we investigated the variation of the first hyperpolarizability in two different fields and plotted the scan. Our simulation results show that mCherry can produce a linear photoelectric effect (the nonzero external wavelength is $271.2$ nm) or second harmonic wave (both external field wavelengths are $271.2$ nm) by adjusting the appropriate external fields, which has great application value in the electro-optic regulator, frequency multiplier, and second-harmonic imaging microscopy\cite{doi:10.1021/ja400098b}.

Given the similarity of RFPs, we suggest that the transition mechanism and optical properties of mCherry can be extricated to other RFPs. The analytical techniques, such as hole-electron analysis and unit spherical representation method used herein, can also be applied to other FPs. At present, there are few detailed analyses of how electrons are transferred in the photoinduced reaction of FPs. The hole-electron analysis is more intuitive and detailed than the electron density method\cite{Nasu2021,Trachman2019} commonly used to analyze FPs and can be used to study electron transfer in various photoinduced reactions. Finally, we believe that our study can serve as a reference for mCherry and other fluorescence protein-related experiments and simulations.

%%%%%%%%%%%%%%%%%%%%%%%%%%%%%%%%%%%%%%%%%%%%%%%%%%%%%%%%%%%%%%%%%%%%%
%% The "Acknowledgement" section can be given in all manuscript
%% classes.  This should be given within the "acknowledgement"
%% environment, which will make the correct section or running title.
%%%%%%%%%%%%%%%%%%%%%%%%%%%%%%%%%%%%%%%%%%%%%%%%%%%%%%%%%%%%%%%%%%%%%
\begin{acknowledgement}

The authors thank financial support from the National Natural Science Foundation of China (12204040). 

\end{acknowledgement}

%%%%%%%%%%%%%%%%%%%%%%%%%%%%%%%%%%%%%%%%%%%%%%%%%%%%%%%%%%%%%%%%%%%%%
%% The same is true for Supporting Information, which should use the
%% suppinfo environment.
%%%%%%%%%%%%%%%%%%%%%%%%%%%%%%%%%%%%%%%%%%%%%%%%%%%%%%%%%%%%%%%%%%%%%

%%%%%%%%%%%%%%%%%%%%%%%%%%%%%%%%%%%%%%%%%%%%%%%%%%%%%%%%%%%%%%%%%%%%%
%% The appropriate \bibliography command should be placed here.
%% Notice that the class file automatically sets \bibliographystyle
%% and also names the section correctly.
%%%%%%%%%%%%%%%%%%%%%%%%%%%%%%%%%%%%%%%%%%%%%%%%%%%%%%%%%%%%%%%%%%%%%
\bibliography{achemso-demo}

\end{document}